\documentclass[prd,aps,preprint,amsmath,amssymb,eshowkeys,nofootinbib]{revtex4-2}
\usepackage[mathscr]{euscript}
\usepackage{dcolumn}
\usepackage{multirow}
\usepackage{bm}
\usepackage{graphicx}
\usepackage{subfigure}
\usepackage{rotating}
\usepackage{amsmath,amssymb,amsthm}
\usepackage[colorlinks=true,linkcolor=red,urlcolor=green,citecolor=green]{hyperref}
\newcommand{\bea}{\begin{eqnarray}}
\newcommand{\eea}{\end{eqnarray}}
\newcommand{\beq}{\begin{equation}}
\newcommand{\eeq}{\end{equation}}

\def\/{\over}

\begin{document}

\title{Growth of structure in holographic dark energy models: a posterior predictive comparison}
\author{Qihong Huang$^{1}$\footnote{Corresponding author: huangqihongzynu@163.com}, Bing Xu$^{2}$, Kaituo Zhang$^{3}$, and Hao Chen$^{1}$}

\affiliation{
$^1$ School of Physics and Electronic Science, Zunyi Normal University, Zunyi, Guizhou 563006, China\\
$^2$ School of Electrical and Electronic Engineering, Anhui Science and Technology University, Bengbu, Anhui 233030, China\\
$^3$ Department of Physics, Anhui Normal University, Wuhu, Anhui 241000, China
}

\begin{abstract}
Structure formation has been used to test the viability of the IR cutoff in the Tsallis holographic dark energy (THDE) model, and the future event horizon was found to be a viable IR cutoff~\cite{Das2026}. Motivated by this, we constrain the HDE, THDE, and BHDE models using the combined Pantheon+SH0ES SN Ia, OHD, DESI DR2, and CMB distance prior datasets, and propagate the MCMC posterior samples into the linear growth sector to obtain the predicted $f\sigma_8(z)$ and $S_8(z)$ histories under the smooth dark energy approximation. We find that the growth predictions of HDE and BHDE are nearly indistinguishable, while those of THDE lie lowest. A posterior predictive $\chi^2$ comparison shows that current $f\sigma_8$ measurements are statistically consistent with all three models, whereas the $S_8$ comparison yields markedly larger $\chi^2$ values for HDE and BHDE and a value for THDE close to that of $\Lambda$CDM. As an illustrative diagnostic, we construct the $(f\sigma_8, S_8)$ plane and find that the trajectories of THDE and $\Lambda$CDM reach the vicinity of the current data region, while those of HDE and BHDE do not. Our results indicate that structure growth observables provide a promising probe for discriminating among holographic dark energy models, while a definitive discrimination will require a joint background and growth likelihood and consistent growth measurements. 
\end{abstract}

\maketitle

\section{Introduction}

Multiple cosmological observational datasets consistently reveal that the expansion of the universe is accelerating~\cite{Perlmutter1999, Riess1998, Spergel2003, Spergel2007, Tegmark2004, Eisenstein2005}. Dark energy offers a natural explanation for this observational fact~\cite{Peebles2003}, with the cosmological constant serving as the simplest and most observationally successful candidate~\cite{Planck2020, ACT2025}. Despite its observational success, the cosmological constant model suffers from several theoretical challenges, including the coincidence problem and the cosmological constant problem~\cite{Weinberg1989, Steinhardt1999}. Both problems arise from our current incomplete understanding of the fundamental origin of the cosmological constant. Recent observations have further exposed significant tensions in measurements of the Hubble constant $H_{0}$ and the matter fluctuation amplitude $\sigma_{8}$~\cite{Planck2020, Riess2022, Abbott2022}, the latter often expressed as the $S_{8}$ degeneracy in weak lensing analyses, raising the possibility of new physics beyond the standard cosmological model~\cite{Valentino2025}. This provides a strong motivation for investigating alternative dark energy models.

As an alternative to the standard dark energy paradigm, holographic dark energy (HDE) is constructed from the holographic principle~\cite{Cohen1999, Hsu2004, Horvat2004, Li2004} and is grounded in the fundamental ideas of quantum gravity. The holographic principle, originating from black hole thermodynamics, suggests that the degrees of freedom in a spatial region are proportional to its boundary area rather than its volume. This principle reflects a deep connection between ultraviolet and infrared physics, commonly referred to as the UV/IR correspondence. In the context of cosmology, this idea can be formulated in an effective quantum field theory, where the zero--point energy density of quantum fluctuations must be linked to a large--scale parameter of the system~\cite{Cohen1999}. The total energy should not exceed the energy required to form a black hole of the same size. This imposes an upper bound on the dark energy density, $L^{3}\rho_{de}\leq L M^{2}_{p}$. By saturating this bound and introducing a dimensionless parameter $C$, the density of HDE is obtained~\cite{Li2004}. Since its proposal, HDE has received extensive interest~\cite{Wang2017}, and is supported by the observational constraints~\cite{Zhang2005, Li2009, Li2025d, Li2025e, Shen2025, Shen2026, Naik2026}. 

Within the HDE framework, different choices of horizon entropy yield different energy densities, thereby classifying HDE into distinct models~\cite{Wang2017}. The original HDE employs the Bekenstein--Hawking entropy~\cite{Cohen1999, Hsu2004, Li2004}; the Tsallis holographic dark energy (THDE) adopts Tsallis entropy, a generalization of the Boltzmann--Gibbs entropy~\cite{Tavayef2018}; and the Barrow holographic dark energy (BHDE) incorporates Barrow's modification of the Bekenstein--Hawking entropy with quantum gravitational corrections from a fractal horizon~\cite{Saridakis2020}. Since their proposal, observational constraints on the parameters of HDE models have been extensively studied. Recently, the Dark Energy Spectroscopic Instrument (DESI) collaboration has released its DR2 BAO measurements~\cite{Abdul2025}, offering the largest spectroscopic galaxy sample to date. This dataset provides a standard ruler for probing cosmic expansion and is now widely applied to constrain dark energy dynamics~\cite{Li2026a, Li2026b, Li2026c, Li2026d, Du2026, Ren2026, Wang2026a, Wang2026b, Du2025, Li2024d, Pang2025, Wang2025a, Yadav2026, Zhu2026, Plaza2025, Petri2026, Li2026, Li2025, Li2025a, Li2025b, Luciano2026, Wang2026, HuangQ2026a, HuangQ2026b, HuangQ2026c, Nagpal2025, Chaudhary2025, Chaudhary2025a, Chaudhary2026, Sharma2026, Capozziello2026, Mehrotra2026, Chaudhary2026a, Chaudhary2026b, Sharma2026a, Chaudhary2026c, Capozziello2026a, Sharma2026b, Chaudhary2027, HuangQ2025d, WangJ2025}. Within the HDE framework, DESI DR2 BAO measurements have also been applied to place observational constraints on the model parameters. With the future event horizon as the IR cutoff, the parameters of HDE have been constrained by the DESI DR2 BAO data combined with various other cosmological observations~\cite{Li2025d, Li2025e, Shen2025, Naik2026}; those of THDE have been constrained by DESI DR2 BAO with CMB and Pantheon+SH0ES SN Ia~\cite{Li2025e}; and the parameter space of BHDE has been restricted by DESI DR2 BAO together with OHD and Pantheon+SH0ES SN Ia~\cite{Luciano2026}. Nevertheless, although the parameters of these models are now constrained by observations, these observational constraints do not yet allow us to distinguish between the original, Tsallis, and Barrow HDE models.

Recently, considering THDE with the particle horizon and the future event horizon as the IR cutoff, by fixing the Tsallis parameter $\delta$ to several representative values and comparing the predictions of $f\sigma_8$ with observational data from redshift--space distortion measurements, it was found that the future event horizon serves as a viable IR cutoff, while the particle horizon is strongly disfavored~\cite{Das2026}. Although a complete assessment of the model requires a full parameter inference using observational data, the result based on fixing the Tsallis parameter $\delta$ to representative values supports the use of the future event horizon as a viable IR cutoff in THDE. This motivates a systematic comparison of HDE, THDE, and BHDE models adopting the future event horizon as the IR cutoff, under a full parameter inference with observational data.

Although parameter constraints for HDE, THDE, and BHDE models have been obtained, they are derived from different data combinations~\cite{Li2025d, Li2025e, Shen2025, Naik2026, Luciano2026}. Nevertheless, the theoretical predictions of $f\sigma_{8}$ and $S_{8}$ depend on $H_{0}$, $\Omega_{m}$, and the model specific parameters, which are constrained from the observational data. To obtain these derived quantities under a consistent observational framework, we re--constrain these models using the same dataset, comprising Pantheon+SH0ES SN Ia, OHD, BAO, and CMB distance priors. Our growth analysis is a posterior predictive comparison: the growth observables are not included in the likelihood used to infer the cosmological parameters; instead, the growth histories are derived from the background only posterior and compared with external measurements a posteriori. Based on the posterior distributions of the model parameters for these HDE models, we then perform this comparison of the predicted $f\sigma_8(z)$ and $S_8(z)$ histories with external growth measurements, in order to assess the current discriminating power of structure formation data among these models.

This paper is organized as follows. In Section~\ref{sec:2}, we briefly review the HDE, THDE, and BHDE models. In Section~\ref{sec:3}, we obtain the observational constraints on their parameters by performing a Markov chain Monte Carlo (MCMC) analysis on the combined Pantheon+SH0ES SN Ia, OHD, DESI DR2, and CMB distance prior dataset. In Section~\ref{sec:4}, we perform a posterior predictive comparison of the linear growth of structure through the $f\sigma_{8}$ and $S_{8}$ observables in these HDE models. Our main conclusions are summarized in Section~\ref{sec:5}.

\section{Models} \label{sec:2}

To explore the HDE models, we adopt a homogeneous and isotropic Friedmann--Robertson--Walker universe with the metric given by
\beq
ds^{2}=-dt^{2}+a^{2}(t)(dr^{2}+r^{2}d\Omega^{2}),
\eeq
and the Friedmann equation is expressed as
\beq
H^{2}=\frac{\kappa^{2}}{3}\big(\rho_{m}+\rho_{de}\big),\label{H2}
\eeq
with $\kappa^{2}=1/M^{2}_{Pl}$, $\rho_{m}$ and $\rho_{de}$ being the energy densities of pressureless matter and dark energy, respectively, and satisfying the conservation equations
\bea
&& \dot{\rho}_{m}+3H\rho_{m}=0,\label{rhom}\\
&& \dot{\rho}_{de}+3H(1+\omega_{de})\rho_{de}=0,\label{rhode}
\eea
where the equation of state parameter $\omega_{de}$ is defined as $\omega_{de}=p_{de}/\rho_{de}$. With the following dimensionless variables introduced
\beq
\Omega_{m}=\frac{\kappa^{2}\rho_{m}}{3H^{2}}, \quad \Omega_{de}=\frac{\kappa^{2}\rho_{de}}{3H^{2}},
\eeq
the Friedmann equation~(\ref{H2}) can be expressed as
\beq
\Omega_{m}+\Omega_{de}=1.\label{O1}
\eeq
and the normalized Hubble parameter $E(z)$ is defined as~\cite{Naik2026}
\beq
E^{2}(z)=\frac{H^{2}(z)}{H^{2}_{0}}=\frac{\Omega_{m0}(1+z)^{3}}{1-\Omega_{de}}.\label{E22}
\eeq
Combining Eqs.~(\ref{H2})--(\ref{rhode}) and~(\ref{O1}), we obtain
\beq
\frac{\dot{H}}{H^{2}}=-\frac{3}{2}(1+\omega_{de}\Omega_{de}).\label{HH2}
\eeq
Employing Eqs.~(\ref{rhode}) and~(\ref{HH2}) along with the definition $'=\frac{d}{dz}$ gives
\beq
\Omega'_{de}=3\omega_{de}(1-\Omega_{de})\Omega_{de}\frac{1}{1+z}.\label{Omde}
\eeq

\subsection{HDE}

When the Bekenstein–-Hawking entropy is applied to the holographic bound on cosmological scales, the HDE model emerges, with the dark energy density given by~\cite{Li2004}
\beq
\rho_{de}=3 C^{2} L^{-2}.\label{rhoHDE0}
\eeq
In this model, an accelerating expansion of the universe is achieved by choosing the future event horizon as the IR cutoff, defined as
\beq
R_{F}=a\int^{\infty}_{a}\frac{da}{H a^{2}},\label{RF}
\eeq
with
\beq
\dot{R}_{F}=H R_{F}-1,\label{RF1}
\eeq
and the dark energy density~(\ref{rhoHDE0}) is expressed as
\beq
\rho_{de}=3 C^{2} R^{-2}_{F}.\label{rhoHDE}
\eeq
Combining Eqs.~(\ref{rhode}),~(\ref{RF1}), and~(\ref{rhoHDE}), we obtain
\beq
\omega_{de}=-\frac{1}{3}-\frac{2\sqrt{\Omega_{de}}}{3C}.
\eeq

\subsection{THDE}

The Tsallis entropy, which is a non-additive generalization of the Boltzmann--Gibbs entropy, has been proposed as~\cite{Tsallis2013}
\beq
S_{\delta}=\gamma A^{\delta},
\eeq
where $\gamma$ is an unknown constant and $\delta$ denotes the non-additivity parameter. The Bekenstein-–Hawking entropy is recovered in the limit $\delta\rightarrow 1$. Based on the Tsallis entropy, THDE model is proposed with the energy density~\cite{Tavayef2018}
\beq
\rho_{de}=B L^{2\delta-4},
\eeq
where $B=3C^{2}M^{2}_{P}$. For $\delta=1$, the standard HDE is recovered. In THDE model, the late--time accelerated expansion of the universe can be realized with either the future event horizon~\cite{Saridakis2018} or the Hubble horizon~\cite{Tavayef2018, Zadeh2018, Huang2019} as the IR cutoff. When the future event horizon is adopted as the IR cutoff, we can express the energy density as follows
\beq
\rho_{de}=B R^{2\delta-4}_{F}.\label{rhoTHDE}
\eeq
Combining Eqs.~(\ref{rhode}),~(\ref{RF1}), and~(\ref{rhoTHDE}) yields
\beq
\omega_{de}=-1-\frac{2\delta-4}{3}\Big(1-\frac{1}{H R_{F}}\Big).
\eeq

\subsection{BHDE}

When quantum--gravitational effects are taken into consideration, it is found that these effects may cause deformations on the black hole surface and lead to deviations from the standard Bekenstein--Hawking entropy~\cite{Barrow2020, Barrow2021}. This modified entropy, known as Barrow entropy, is defined as
\beq
S_{B}=\Big(\frac{A}{A_{0}}\Big)^{1+\frac{\Delta}{2}},
\eeq
where $A$ denotes the standard horizon area, $A_{0}$ corresponds to the Planck area, and $\Delta$ is a parameter quantifies quantum--gravitational deformation, with the range $0 \leq \Delta \leq 1$. The limits $\Delta=0$ and $\Delta=1$ correspond to a smooth spacetime and the most intricate deformation, respectively. When the Barrow entropy is applied to the UV/IR relation, the BHDE model is proposed~\cite{Saridakis2020}, with the energy density
\beq
\rho_{de}=B L^{\Delta-2},
\eeq
where $B=3C^{2}M^{2}_{P}$. For $\Delta=0$, the standard HDE is recovered. Similar to the THDE case, the BHDE model also allows for late--time accelerated expansion when the IR cutoff is taken as either the future event horizon~\cite{Saridakis2020, Huang2021} or the Hubble horizon~\cite{Huang2021}. With the future event horizon as the IR cutoff, the energy density is given by
\beq
\rho_{de}=B R^{\Delta-2}_{F}.\label{rhoBHDE}
\eeq
Combining Eqs.~(\ref{rhode}),~(\ref{RF1}), and~(\ref{rhoBHDE}), we get
\beq
\omega_{de}=-1-\frac{\Delta-2}{3}\Big(1-\frac{1}{H R_{F}}\Big).
\eeq

\section{Observational Analysis} \label{sec:3}

In previous section, we briefly review HDE, THDE, and BHDE models. In this section, we estimate the model parameters using the recent Pantheon+SH0ES SN Ia sample~\cite{Riess2022, Brout2022, Scolnic2022}, the observational Hubble data (OHD)~\cite{Cao2022}, DESI DR2 BAO data~\cite{Abdul2025}, and Planck 2018 CMB distance priors~\cite{Zhai2019,Chen2019}. After briefly introducing the datasets and the Bayesian methodology, we apply them to these HDE models and present the resulting observational constraints.

\subsection{Data and Methodology}

\subsubsection{Pantheon+SH0ES SN Ia Sample}

The observation of high--redshift type Ia supernovae has remained a critical probe of the late--time universe since the initial discovery of accelerated expansion. This probe involves comparing the observed SN Ia luminosity distance with the theoretical predictions of a given model. The Pantheon+SH0ES SN Ia sample is currently the largest high--redshift SN Ia compilation, covering the redshift range $z \in [0.001, 2.261]$~\cite{Riess2022, Brout2022, Scolnic2022}. Because the redshifts of Pantheon+SH0ES SN Ia below $z = 0.01$ are strongly affected by unmodeled peculiar velocities~\cite{Brout2022}, we exclude them from the SN sample. For the SN Ia sample, the $\chi^{2}$ term is defined as 
\beq
\chi^{2}_{SN}=\left( \hat{m}_{obs} - m_{th} \right)^\dag C_{SN}^{-1} \left( \hat{m}_{obs} - m_{th} \right),
\eeq
where $\hat{m}_{obs}$ is the vector of observed corrected apparent magnitude and $C_{SN}$ denotes the corresponding covariance matrix. The predicted apparent magnitude $m_{th}$ takes the form
\beq
m_{th}=5 \log_{10} \Big( \frac{D_{L}(z)}{Mpc} \Big) + 25 + M,
\eeq
where $M$ denotes the absolute magnitude of SN Ia and $D_{L}(z)$ is the luminosity distance, defined as
\beq
D_{L}(z)=c(1+z)\int^{z}_{0}\frac{dz}{H(z)}.
\eeq

\subsubsection{Observational Hubble Data}

The Hubble parameter $H(z)$, which characterizes the expansion rate of the universe as a function of redshift $z$, is currently constrained by baryon acoustic oscillation measurements and by the differential age of passive galaxies, a method often referred to as cosmic chronometers. For the OHD sample, the $\chi^{2}$ term takes the form
\beq
\chi^{2}_{OHD} = \sum_{i=1}^{N_{OHD}} \left(\frac{H_{obs,i}-H_{th}(z_i)}{\sigma_{OHD,i}}\right)^2
\eeq
where $H_{obs,i}$ denotes the $i$-th observed value and $\sigma_{OHD,i}$ its corresponding standard deviation, with $N_{OHD}$ being the total number of OHD data. 

\subsubsection{DESI DR2 BAO}

BAOs are density fluctuations imprinted in the matter distribution by acoustic oscillations in the pre--recombination plasma, and have been accurately measured by DESI DR2 based on luminous red galaxies over the redshift range $z \in [0.295,2.33]$~\cite{Abdul2025}. For the DESI DR2 BAO sample, the $\chi^{2}$ term is given by
\beq
\chi^{2}_{DESI} = \sum_{i=1}^{N_{DESI}} \Delta_{i}^{T} \mathbf{C}_{i}^{-1} \Delta_{i},
\eeq
with $\Delta_{i}$ and $\mathbf{C}_{i}$ depending on the measurement type in the $i$-th redshift bin. For anisotropic measurements, we compute them as
\bea
\Delta_{i}=
\left(
\begin{matrix}
D_{M}^{obs}(z_{i})/r_{d}-D_{M}^{th}(z_{i})/r_{d}\\[6pt]
D_{H}^{obs}(z_{i})/r_{d}-D_{H}^{th}(z_{i})/r_{d}
\end{matrix}
\right)
\eea
and
\bea
\mathbf{C}_{i}=
\left(
\begin{matrix}
\sigma^{2}_{D_M/r_d} & r_{HM} \sigma_{D_M/r_d} \sigma_{D_H/r_d}\\
r_{HM} \sigma_{D_M/r_d} \sigma_{D_H/r_d} & \sigma^{2}_{D_H/r_d}
\end{matrix}
\right).
\eea
When isotropic measurements are considered, $\Delta_{i}$ reduces to the scalar $D_{V}^{obs}(z_{i})/r_{d}-D_{V}^{th}(z_{i})/r_{d}$, with variance $\sigma^{2}_{D_V/r_d}$. The Hubble distance and the transverse comoving distance are given by
\beq
D_{H}(z)=\frac{c}{H(z)},
\eeq
and
\beq
D_{M}(z)=c\int^{z}_{0}\frac{dz}{H(z)},
\eeq
and the volume averaged distance $D_{V}(z)$ in the isotropic limit takes the form
\beq
D_{V}(z)=\left[ z D_{M}^{2}(z) D_{H}(z) \right]^{1/3}.
\eeq
Here, $\sigma_{D_{M}/r_{d}}$ and $\sigma_{D_{H}/r_{d}}$ are the uncertainties for $D_{M}(z)/r_{d}$ and $D_{H}(z)/r_{d}$, $r_{HM}$ is their correlation coefficient, and $r_{d}$ is the sound horizon at drag.

\subsubsection{CMB Distance Priors}

CMB distance priors are derived from the Planck 2018 measurements based on the cosmic microwave background anisotropies, providing constraints on the CMB shift parameter $R$, the acoustic scale $l_{A}$, and the baryon density $\Omega_{b}h^{2}$~\cite{Zhai2019,Chen2019}. For the CMB distance priors, the $\chi^{2}$ term is computed as
\beq
\chi^{2}_{CMB} = \Delta p^{T} \mathbf{C}_{CMB}^{-1} \Delta p, \qquad \Delta p=p^{obs}-p^{th},
\eeq
where $p=\{R,l_{A},\Omega_{b}h^{2}\}$, $\mathbf{C}_{CMB}$ is the covariance matrix, and $R$ and $l_{A}$ are given by
\beq
R=\frac{D_{M}(z_{*})\sqrt{\Omega_{m}H^{2}_{0}}}{c}, \qquad l_{A}=\frac{\pi D_{M}(z_{*})}{r_{s}(z_{*})},
\eeq
where $z_{*}$ denotes the redshift at photon decoupling. From Planck CMB observations, $p^{\text{obs}}=\{1.7502, 301.471, 0.02236\}$. Given the corrections between $\Omega_{b}h^{2}$, $R$, and $l_{A}$, their covariance matrix from~\cite{Chen2019} is included in the $\chi^{2}$ calculation.

\subsubsection{Methodology}

To estimate the free parameters for these HDE models, we perform a Bayesian analysis using the MCMC sampling with the emcee library in Python. The burn--in is set to twice the maximum integrated autocorrelation time, and convergence is verified by requiring that the chain length be at least 50 times the autocorrelation time for each parameter. For a direct comparison with the standard $\Lambda$CDM model, we carry out an additional MCMC analysis for $\Lambda$CDM using the same datasets. For the Pantheon+SH0ES SN Ia, OHD, and DESI DR2 BAO datasets, the prior ranges for the free parameters of these HDE models are listed in Table~\ref{Tab1}; in this data combination, $r_d$ is treated as an additional free parameter, with prior range $[130, 160]$. For the Pantheon+SH0ES SN Ia, OHD, DESI DR2 BAO datasets, and Planck 2018 CMB distance priors datasets, the prior ranges for the free parameter $B$ is set to [0.0,4.0], $r_{d}$ is computed from the standard integration formula using the Planck 2018 best--fit drag redshift $z_{drag}=1059.94$, while $r_{*}$, the sound horizon at photon decoupling, is obtained from the standard integration formula using the Planck 2018 best--fit value $z_{*}=1089.92$~\cite{Planck2020}.

\begin{table}[h]
\centering
\caption{\label{Tab1} Free parameters and prior ranges for the $\Lambda$CDM and HDE models.}
 \begin{tabular}{|c|c|c|c|c|c|c|c|}
  \hline
  \hline
  Model & $H_{0}$ & $\Omega_{m}$ & C & B & $\delta$ & $\Delta$ & M\\
  \hline
  $\Lambda$CDM & [50,80] & [0.1,0.5] & $-$ & $-$ & $-$ & $-$ & [-20,-18]\\
  \hline
  HDE & [50,80] & [0.1,0.5] & [0.0,2.0] & $-$ & $-$ & $-$ & [-20,-18]\\
  \hline
  THDE & [50,80] & [0.1,0.5] & $-$ & [2.0,4.0] & [0.0,2.0] & $-$ & [-20,-18]\\
  \hline
  BHDE & [50,80] & [0.1,0.5] & $-$ & [2.0,4.0] & $-$ & [0.0,2.0] & [-20,-18]\\
  \hline
  \hline
  \end{tabular}
\end{table}

The total log--likelihood is given by
\beq
\ln(\mathcal{L}_{total})=-\frac{1}{2}\chi^{2}_{total}+const.,
\eeq
with
\beq
\chi^{2}_{total}=\chi^{2}_{SN}+\chi^{2}_{OHD}+\chi^{2}_{DESI}+\chi^{2}_{CMB}.
\eeq
For statistical model comparison, we adopt the Akaike Information Criterion (AIC)~\cite{Akaike1974} and Bayesian Information Criterion (BIC)~\cite{Schwarz1978, Trotta2008}, which penalize models with different numbers of parameters. The criteria are defined as
\beq
AIC=\chi^{2}_{min}+2n,
\eeq
and
\beq
BIC=\chi^{2}_{min}+n\ln(N),
\eeq
where $n$ denotes the number of parameters and $N$ the total number of data points. 

\subsection{Results from observational constraints}

\begin{figure}[h]
\begin{center}
\includegraphics[width=0.45\textwidth]{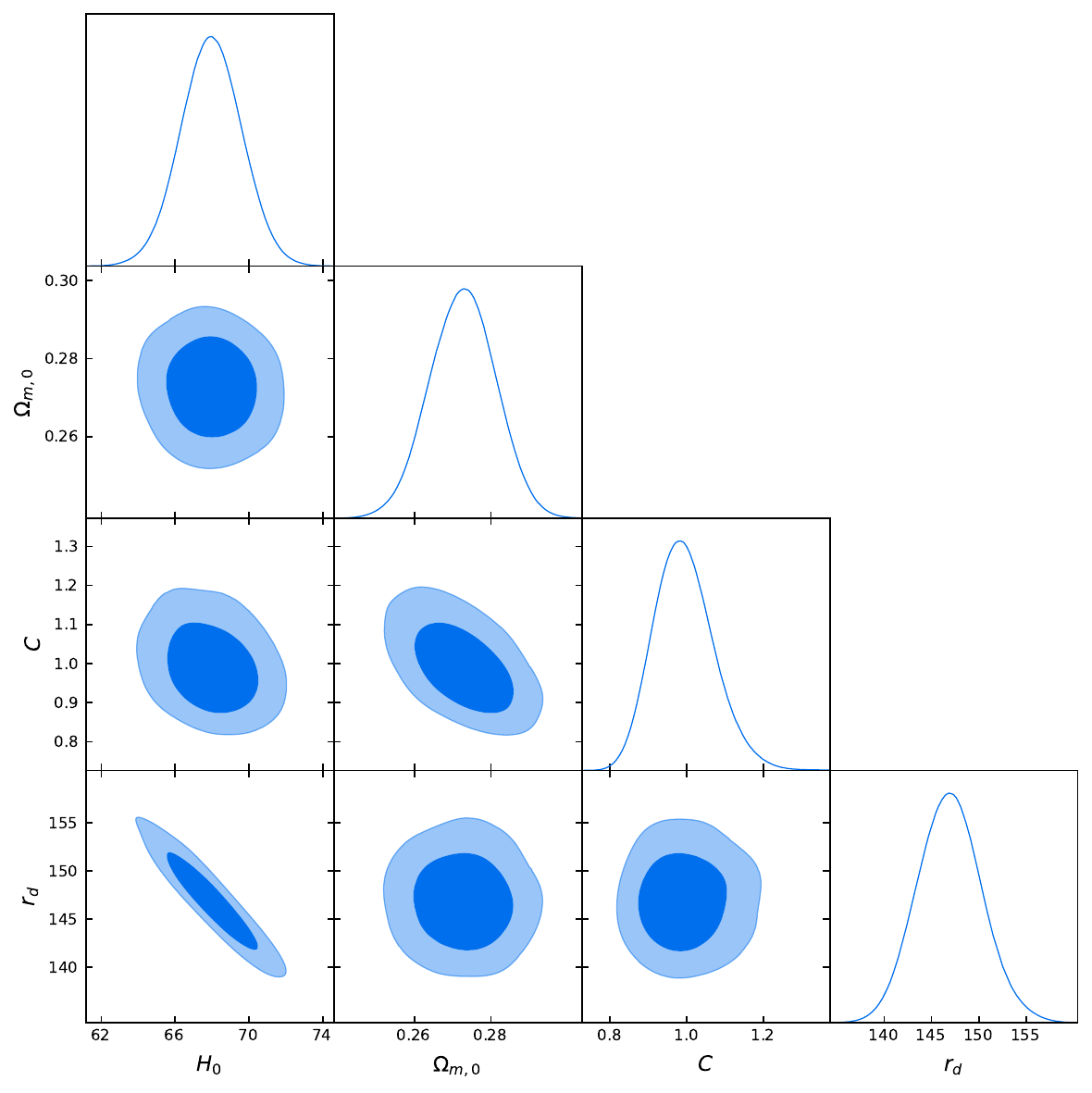}
\includegraphics[width=0.45\textwidth]{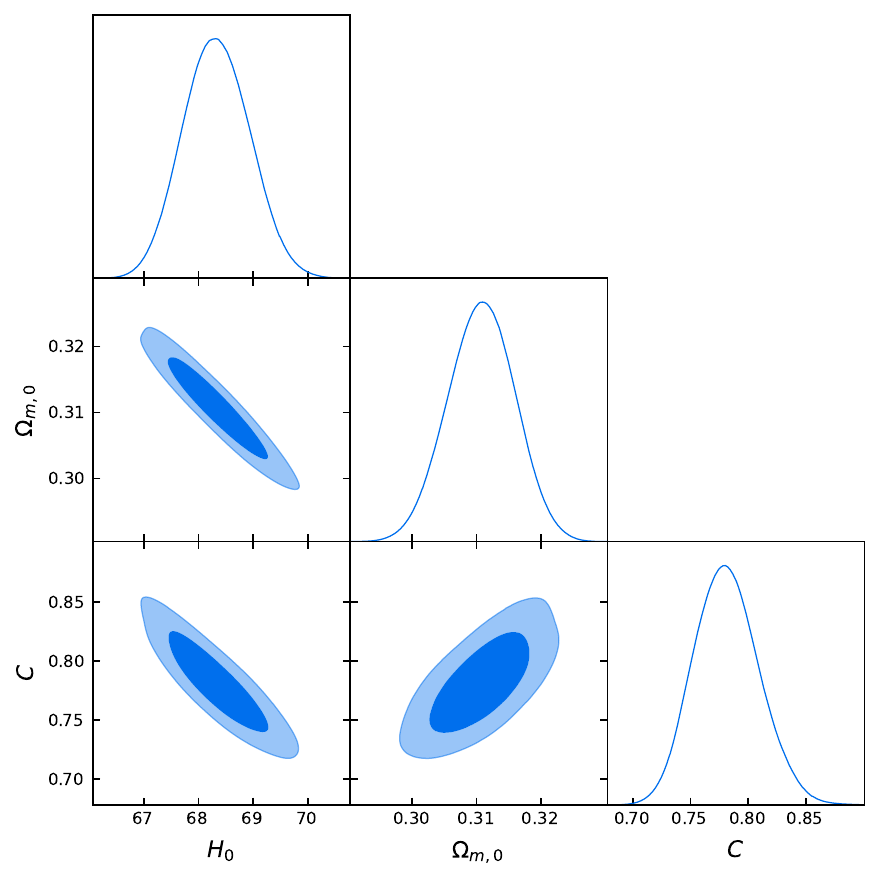}
\caption{\label{Fig1} Posterior distributions for the free parameters of the HDE model. The left panel uses Pantheon+SH0ES SN Ia, OHD, and DESI DR2, while the right panel adds Planck 2018 CMB distance priors.}
\end{center}
\end{figure}

\begin{figure}[h]
\begin{center}
\includegraphics[width=0.45\textwidth]{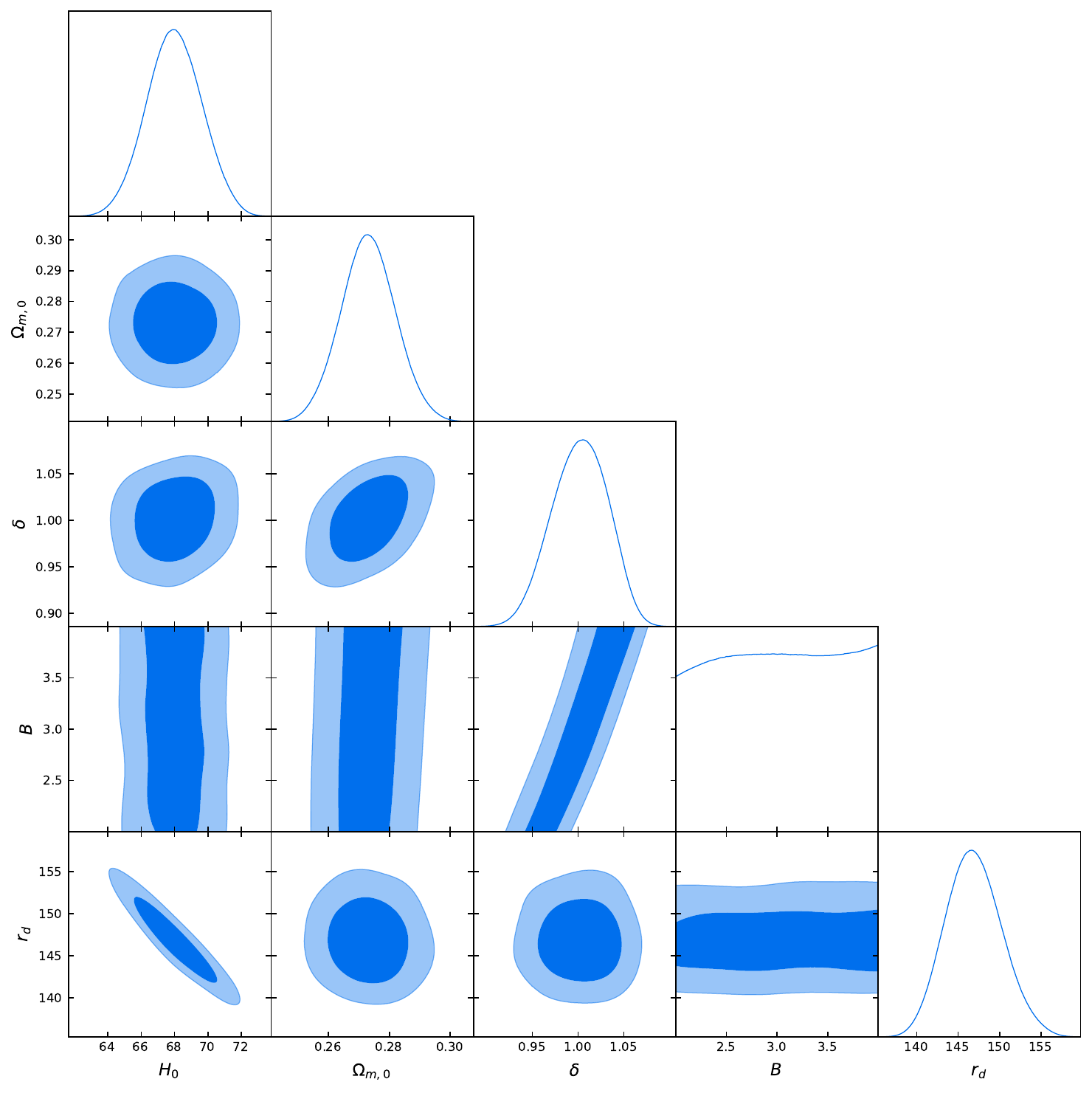}
\includegraphics[width=0.45\textwidth]{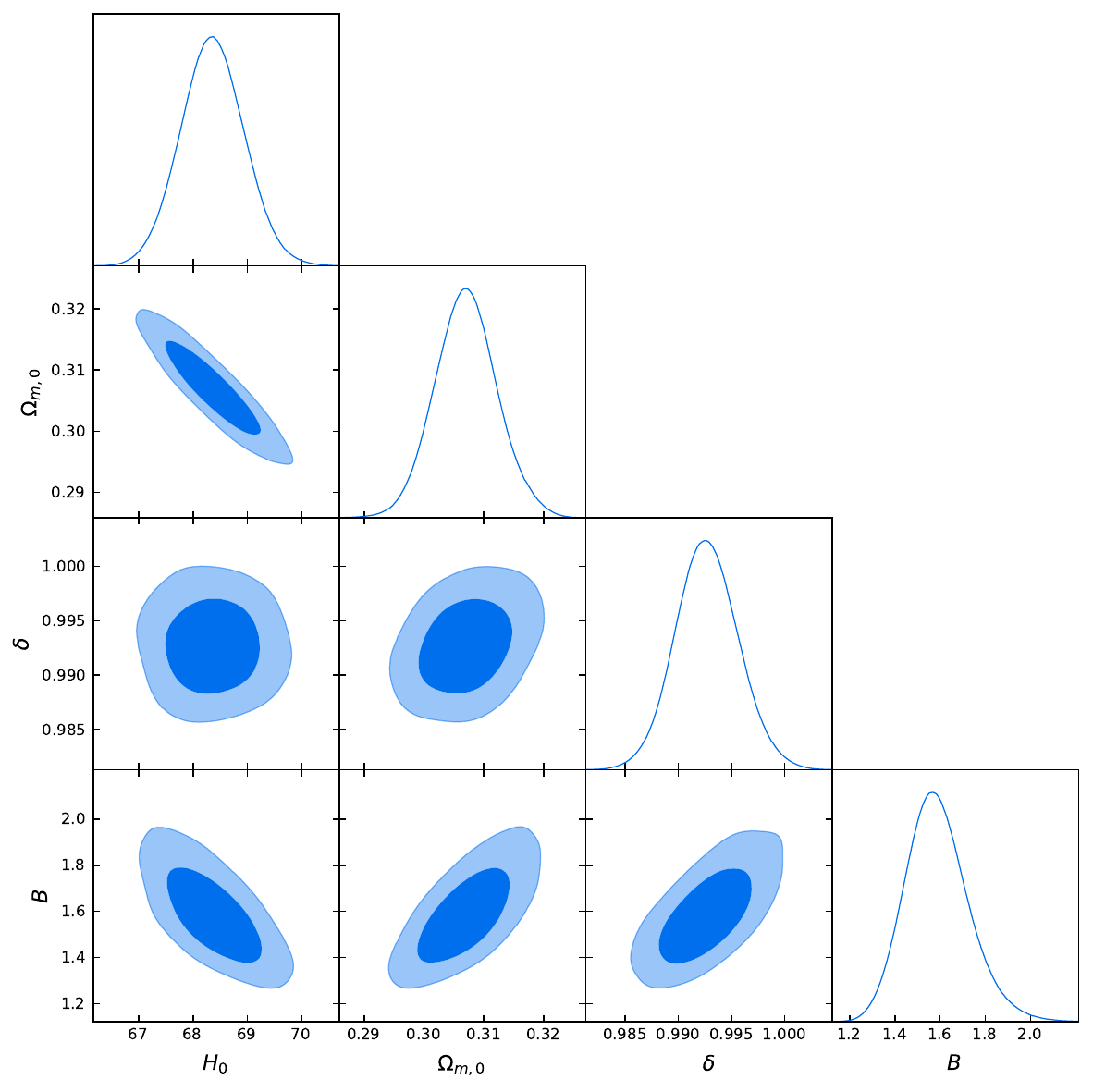}
\caption{\label{Fig2} Posterior distributions for the free parameters of the THDE model. The left panel uses Pantheon+SH0ES SN Ia, OHD, and DESI DR2, while the right panel adds Planck 2018 CMB distance priors.}
\end{center}
\end{figure}

\begin{figure}[h]
\begin{center}
\includegraphics[width=0.45\textwidth]{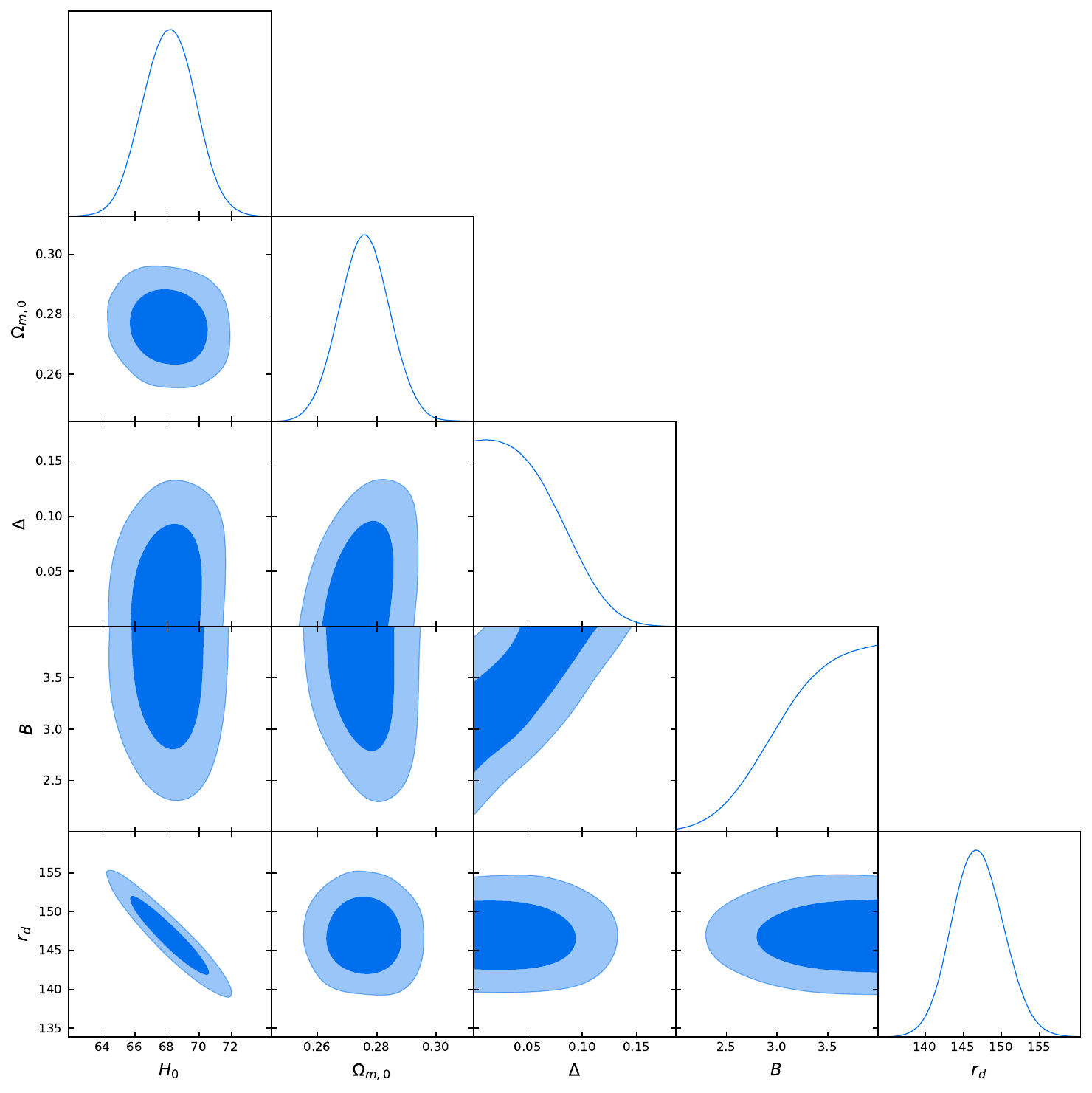}
\includegraphics[width=0.45\textwidth]{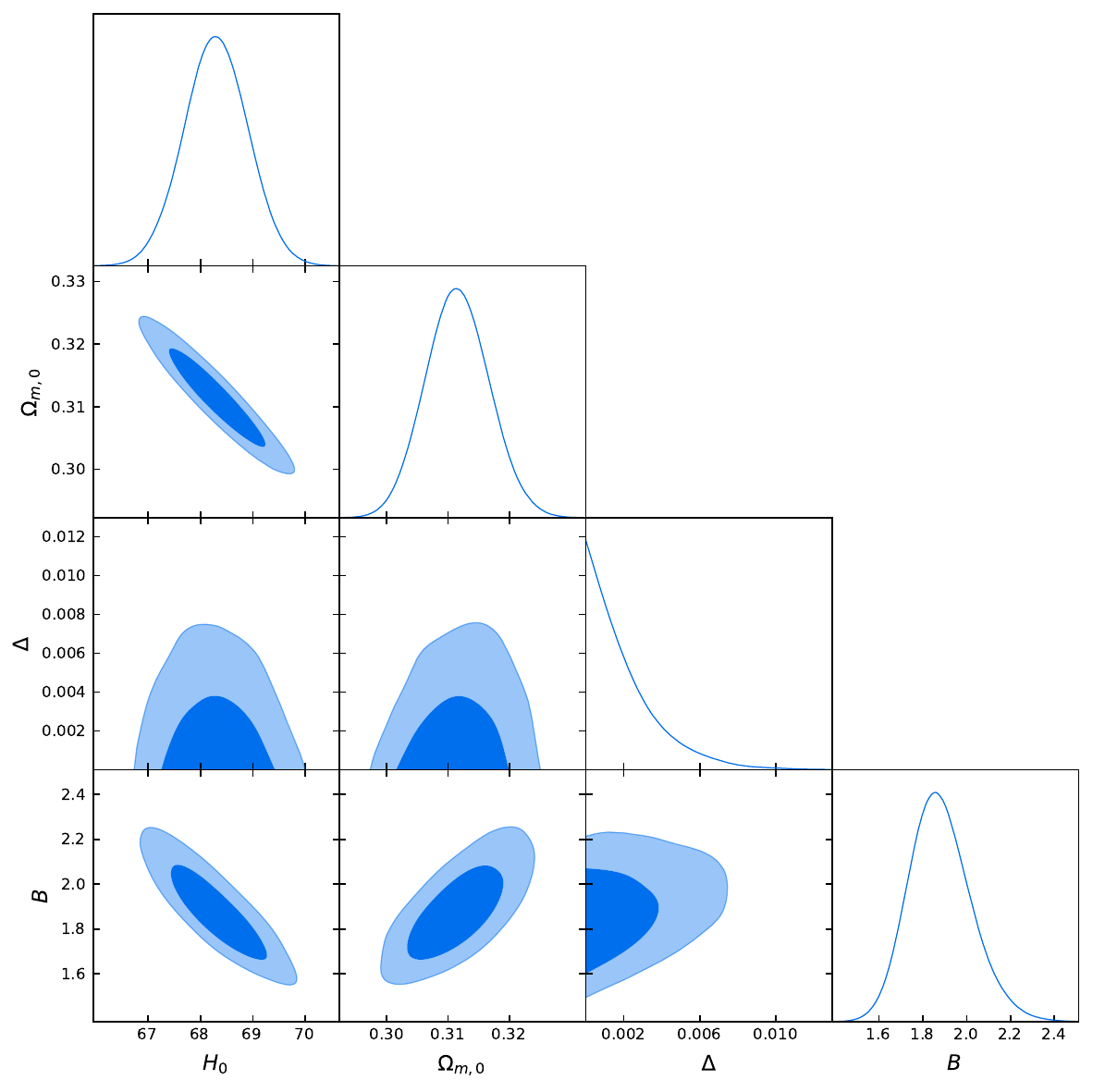}
\caption{\label{Fig3} Posterior distributions for the free parameters of the BHDE model. The left panel uses Pantheon+SH0ES SN Ia, OHD, and DESI DR2, while the right panel adds Planck 2018 CMB distance priors.}
\end{center}
\end{figure}

Table~\ref{Tab2} summarizes the parameter constraints for the $\Lambda$CDM, HDE, THDE, and BHDE models, listing the mean values and 1$\sigma$ confidence levels (CL), where three centered dots ($\cdots$) indicate that the mean values cannot be calculated because the observational constraints are nearly flat across the entire range. The quantities $\Delta\chi^{2}_{min}$, $\Delta AIC$, and $\Delta BIC$ are defined relative to $\Lambda$CDM, with negative values favoring these HDE models and positive values favoring $\Lambda$CDM. The corresponding posterior distributions for the HDE, THDE, and BHDE models are displayed in Figs.~\ref{Fig1}, ~\ref{Fig2}, and ~\ref{Fig3}, respectively.

For the combination of Pantheon+SH0ES SNe Ia, OHD, and DESI DR2 datasets, the sound horizon $r_{d}$ is treated as a free parameter, and its posterior distributions are shown in the left panels of Figs.~\ref{Fig1}, ~\ref{Fig2}, and ~\ref{Fig3}. Specifically, as shown in Table~\ref{Tab2}, the constraints on $r_{d}$ are highly consistent for these models, yielding $r_d \approx 146.9$ Mpc. These values are in good agreement with the Planck 2018 results of $r_d \approx 147.0$ Mpc~\cite{Planck2020}. When the Planck 2018 CMB distance priors are included, $r_{d}$ is instead computed as a derived quantity from the standard integration formula using the Planck 2018 best--fit drag redshift $z_{drag}=1059.94$, rather than being sampled as a free parameter. Because it is a derived quantity in this case, it is not sampled in the MCMC chains and its posterior is therefore not displayed in the right panels of Figs.~\ref{Fig1}, ~\ref{Fig2}, and ~\ref{Fig3}; correspondingly, it is not counted in the number of free parameters $n$ for the AIC and BIC calculations.

\begin{table}[h]
\centering
\caption{\label{Tab2} Observational constraints ($68\%$ CL) on the $\Lambda$CDM, HDE, THDE, and BHDE models from different data combinations.}
\resizebox{\textwidth}{!}{
\begin{tabular}{|c|c|c|c|c|c|c|c|c|c|c|c|c|}
\hline
\hline
Datasets & Model & $H_{0}$ & $\Omega_{m,0}$ & $r_{d}$ & C & B & $\delta$ & $\Delta$ & $\chi^{2}_{min}$ & $\Delta \chi^{2}_{min}$ & $\Delta AIC$ & $\Delta BIC$\\
\hline
\multirow{4}{*}{\shortstack{SN\\+OHD\\+DESI DR2}} & $\Lambda$CDM & $68.74^{+1.62}_{-1.64}$ & $0.305^{+0.007}_{-0.008}$ & $146.99^{+3.21}_{-3.59}$ & $-$ & $-$ & $-$ & $-$ & $1431.4$ & $0$ & $0$ & $0$\\
\cline{2-13}
& HDE & $67.96^{+1.62}_{-1.65}$ & $0.273^{+0.008}_{-0.009}$ & $146.92^{+3.31}_{-3.45}$ & $0.993^{+0.068}_{-0.085}$ & $-$ & $-$ & $-$ & $1428.3$ & $-3.1$ & $-1.1$ & $4.3$\\
\cline{2-13}
& THDE & $67.99^{+1.68}_{-1.66}$ & $0.273\pm 0.009$ & $146.89^{+3.18}_{-3.58}$ & $-$ & $\cdots$ & $1.002^{+0.032}_{-0.031}$ & $-$ & $1428.3$ & $-3.1$ & $0.9$ & $11.6$\\
\cline{2-13}
& BHDE & $68.13^{+1.63}_{-1.62}$ & $0.276^{+0.008}_{-0.009}$ & $146.93^{+3.31}_{-3.34}$ & $-$ & $>3.193$ & $-$ & $<0.064$ & $1428.2$ & $-3.2$ & $0.8$ & $11.6$\\
\hline
\multirow{4}{*}{\shortstack{SN\\+OHD\\+DESI DR2\\+CMB}} & $\Lambda$CDM & $68.33^{+0.23}_{-0.24}$ & $0.304^{+0.004}_{-0.003}$ & $-$ & $-$ & $-$ & $-$ & $-$ & $1437.6$ & $0$ & $0$ & $0$\\
\cline{2-13}
& HDE & $68.33 \pm 0.61$ & $0.311 \pm 0.005$ & $-$ & $0.781^{+0.026}_{-0.030}$ & $-$ & $-$ & $-$ & $1456.8$ & $19.2$ & $21.2$ & $26.6$\\
\cline{2-13}
& THDE & $68.36 \pm 0.57$ & $0.307 \pm 0.005$ & $-$ & $-$ & $1.588^{+0.120}_{-0.151}$ & $0.993^{+0.002}_{-0.003}$ & $-$ & $1450.4$ & $12.8$ & $16.8$ & $27.7$\\
\cline{2-13}
& BHDE & $68.30 \pm 0.60$ & $0.312^{+0.005}_{-0.006}$ & $-$ & $-$ & $1.881^{+0.122}_{-0.155}$ & $-$ & $<0.002$ & $1456.9$ & $19.3$ & $23.3$ & $34.1$\\
\hline
\hline
\end{tabular}
}
\end{table}

For the $\Lambda$CDM model, a comparison of the two datasets shows that the CMB distance priors are consistent with the low--redshift data. Including the CMB priors results in only a slight shift in the mean values of $H_{0}$, $\Omega_{m,0}$, and $\chi^{2}_{min}$ for the $\Lambda$CDM model.

With the combined Pantheon+SH0ES SN Ia, OHD, and DESI DR2 data, these HDE models give nearly the same mean values of $H_{0}$, $\Omega_{m,0}$, and $\chi^{2}_{min}$. For the THDE and BHDE models, the mean values of $\chi^{2}_{min}$, $\Delta \chi^{2}_{min}$, $\Delta$AIC, and $\Delta$BIC are very close. However, the parameter $B$ is not constrained for THDE model, whereas for BHDE model it is found to be bounded from below, $B>3.193$. For this data combination, the HDE, THDE, and BHDE models yield marginally lower $\chi^{2}_{min}$ values compared to $\Lambda$CDM. The $\Delta$AIC values for HDE, THDE, and BHDE are $-1.1$, $0.9$, and $0.8$, respectively. According to the AIC, HDE is marginally preferred over $\Lambda$CDM, while THDE and BHDE are slightly disfavored. However, the $\Delta$BIC values for HDE, THDE, and BHDE are $4.3$, $11.6$, and $11.6$, respectively, indicating that these models are increasingly disfavored due to the penalty for the extra parameters $B$, $\delta$, and $\Delta$. 

When the CMB distance priors are included, the mean values of $H_{0}$ and $\Omega_{m,0}$ are nearly same for all these models. Compared to the results obtained from the previous datasets, the inclusion of the CMB distance priors tightens the constraints on all model parameters $C$, $B$, $\delta$, and $\Delta$. However, the values of $\Delta \chi^{2}_{min}$, $\Delta$AIC, and $\Delta$BIC for these HDE models are all larger than $10$, indicating that these HDE models are strongly disfavored relative to the $\Lambda$CDM model.

For the HDE model, constraining the free parameters with the combined CMB, DESI, and Pantheon+SH0ES SN Ia data yields $H_{0}=66.65 \pm 0.63$, $\Omega_{m,0}=0.317 \pm 0.006$, and $C=0.673 \pm 0.023$~\cite{Li2025d}. When the free parameters are constrained with the combined BAO, SN Ia, OHD, $H_{0}$, and CMB distance prior data, one obtains $H_{0}=68.81 \pm 0.48$, $\Omega_{m,0}=0.2911 \pm 0.0042$, and $C=0.711 \pm 0.021$~\cite{Shen2025}. From our analysis of the combined Pantheon+SH0ES SN Ia, OHD, DESI DR2, and CMB distance prior data, we obtain $H_{0}=68.33 \pm 0.61$, $\Omega_{m,0}=0.311 \pm 0.005$, and $C=0.781^{+0.026}_{-0.030}$. The comparison shows that our results are in good agreement with those of~\cite{Shen2025}, while some differences are observed compared to~\cite{Li2025d}, indicating that different data combinations can lead to variations in the inferred cosmological parameters, particularly for $H_{0}$ and $\Omega_{m,0}$.

For the THDE model, fitting the free parameters to the combined cosmic chronometer (CC) and SN Ia data gives $H_{0}=67.02^{+1.03}_{-1.02}$, $\Omega_{m,0}=0.3080^{+0.0451}_{-0.0521}$, and $\delta=1.0709^{+0.0584}_{-0.0648}$~\cite{Saridakis2018}. When the free parameters are constrained with the combined BAO, CMB, Pantheon, Gamma--Ray burst, and local $H_{0}$ data, it is found that $H_{0}=68.851\pm 0.600$, $\Omega_{de,0}=0.695\pm 0.018$, and $\delta=1.071 \pm 0.023$~\cite{Sadri2019}. Our constraints from the combined Pantheon+SH0ES SN Ia, OHD, DESI DR2, and CMB distance prior data are $H_{0}=68.36 \pm 0.57$, $\Omega_{m,0}=0.307 \pm 0.005$, and $\delta=0.993^{+0.002}_{-0.003}$. A comparison of these three sets of constraints reveals that our $H_{0}$ and $\Omega_{m,0}$ results are in good agreement with those of~\cite{Saridakis2018} and~\cite{Sadri2019}, respectively, with significantly improved precision. However, the parameter $\delta$ shows a discrepancy between our result and the other two studies, suggesting that it is sensitive to the choice of data combination.

For the BHDE model, the combined CC and SN Ia data gives $H_{0}=68.92^{+1.87}_{-1.89}$, $\Omega_{m,0}=0.284^{+0.043}_{-0.044}$, and $\Delta=0.094^{+0.094}_{-0.101}$~\cite{Anagnostopoulos2020}, while the combined SN Ia, OHD, and BAO data gives $H_{0}=68.6^{+1.3}_{-3.3}$, $\Omega_{m,0}=0.316^{+0.020}_{-0.023}$, and $\Delta<0.471$~\cite{Luciano2026}. Our analysis of the combined Pantheon+SH0ES SN Ia, OHD, DESI DR2, and CMB distance prior data yields $H_{0}=68.30 \pm 0.60$, $\Omega_{m,0}=0.312^{+0.005}_{-0.006}$, and $\Delta<0.002$. A comparison of these three sets of constraints shows that our $H_{0}$ results are in good agreement with previous studies, while our $\Omega_{m,0}$ constraint is consistent with~\cite{Luciano2026}, but shows a slight deviation from~\cite{Anagnostopoulos2020}. Most notably, our upper limit on $\Delta$ is significantly tighter than those obtained in earlier works, and is consistent with the BBN limit $\Delta<1.4\times 10^{-4}$~\cite{Barrow2021}, suggesting that the BHDE model with updated data favors $\Delta$ approaching $0$, effectively recovering the standard Bekenstein--Hawking entropy.

Overall, across these HDE models, the cosmological parameters $H_{0}$ and $\Omega_{m,0}$ are robust against variations in the data combination, whereas the model parameters exhibit data dependence. When CMB distance priors are included, these HDE models provide fit statistics comparable to $\Lambda$CDM model, with THDE exhibiting the best fit among these HDE models. However, $\Lambda$CDM remains statistically favored overall once the number of free parameters is accounted for, and these HDE models are disfavored.

\subsection{Evolution of the cosmological parameters}

\begin{figure*}[htp]
\begin{center}
\includegraphics[width=0.45\textwidth]{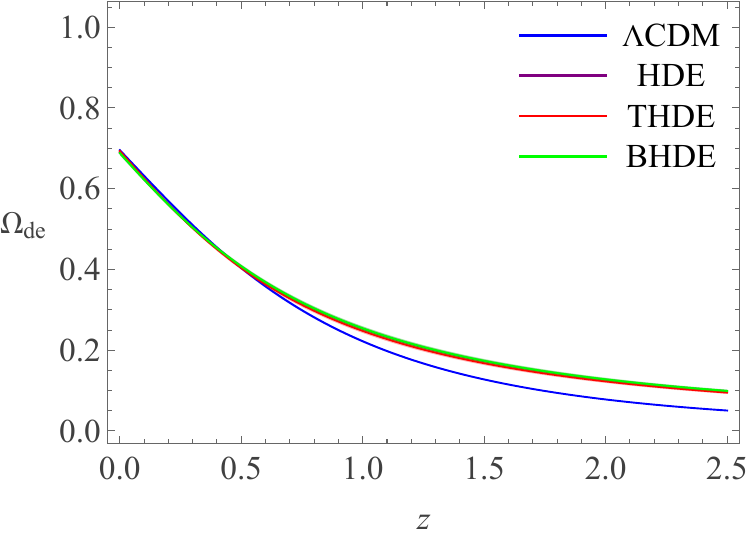}
\includegraphics[width=0.47\textwidth]{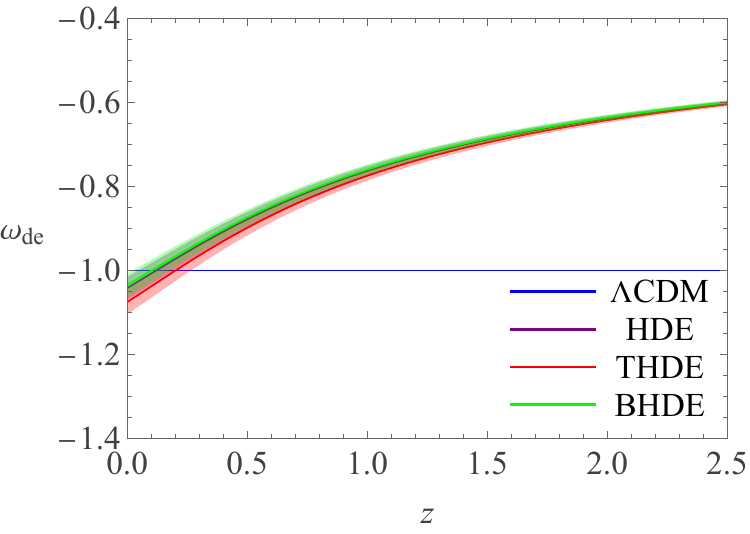}
\caption{\label{Fig4} Evolutionary curves of $\Omega_{de}$ and $\omega_{de}$ for these models with the mean values and 68\% CL regions, obtained from the Pantheon+SH0ES SN Ia, OHD, DESI DR2, and CMB distance prior data.}
\end{center}
\end{figure*}

To analyze the cosmological evolution of $\Omega_{de}$ and $\omega_{de}$ for these models, we solve the dynamical equation~(\ref{Omde}) for each set of parameters extracted from the MCMC samples, and compute the mean values and 68\% CL regions to plot the evolution curves with their 68\% confidence bands. The resulting evolutionary curves are shown in Fig.~\ref{Fig4}, with the $\Lambda$CDM curve plotted as a blue solid line.

The 68\% CL regions of $\Omega_{de}$ and $\omega_{de}$ reveal that HDE and BHDE are indistinguishable in these observables, while THDE is only partially separated from them. As shown in the left panel of Fig.~\ref{Fig4}, $\Omega_{de}$ approaches $0$ at early times and reaches $0.7$ at late times for all models, implying that dark energy dominates the late time cosmic evolution. For HDE and BHDE models, the evolutionary curves of $\Omega_{de}$ overlap completely, while that of THDE deviates from them. All three models deviate significantly from the curves for the $\Lambda$CDM model at early times, but converge to it at late times. The right panel shows that $\omega_{de}$ for HDE, THDE, and BHDE models evolves from the quintessence regime at early times to the phantom regime at late times. The three curves overlap completely at early times, while the THDE model diverges from the others at late times, with HDE and BHDE being nearly indistinguishable. 

Based on the above analysis, the HDE and BHDE models are indistinguishable in the evolution of $\Omega_{de}$, whereas the THDE model is only partially separated from them. The $\omega_{de}$ curves show differences at late times, but HDE and BHDE remain nearly indistinguishable in $\omega_{de}$ as well. Therefore, these observables alone are insufficient to fully distinguish the three models.

\section{Linear Growth of Matter Perturbations} \label{sec:4}

In the previous section, we have constrained the cosmological and model parameters of these HDE models using the combined observational data, analyzed the evolution of $\Omega_{de}$ and $\omega_{de}$ using each set of parameters extracted from the MCMC samples, and found that these models are difficult to distinguish using the evolution of $\Omega_{de}$ and $\omega_{de}$ alone. Recently, a viable IR cutoff in THDE has been identified by comparing the predictions of $f\sigma_8$ with observational data from redshift--space distortion measurements~\cite{Das2026}. In this section, we apply the same strategy to the HDE, THDE, and BHDE models, using the parameter sets drawn from the MCMC samples of the Pantheon+SH0ES SN Ia, OHD, DESI DR2, and CMB distance prior data. This is a posterior predictive comparison conditioned on the background only posterior: the growth measurements are not part of the likelihood, and the comparison below therefore assesses the consistency of the background constrained models with external growth data, rather than constituting a joint background and growth inference.

The evolution of large--scale structure, driven by the growth of matter density perturbations, provides a probe of dark energy models that is complementary to background expansion probes. Under the assumptions of negligible dark energy perturbations on subhorizon scales and the standard smooth dark energy approximation, the growth of structure is governed solely by matter perturbations within general relativity~\cite{Das2026}. With pressureless matter and negligible anisotropic stress, the subhorizon evolution of the matter density contrast $\delta$ is described by the linear perturbation equation
\beq
\ddot{\delta}+2H \dot{\delta}-4\pi G \rho_{m}\delta =0,\label{delta2}
\eeq
where $\delta=\delta\rho_{m}/\rho_{m}$. Taking the scale factor $a$ as the independent variable, Eq.~(\ref{delta2}) becomes
\beq
\delta''(a)+\Big[\frac{H'(a)}{H(a)}+\frac{3}{a}\Big]\delta'(a)-\frac{3}{2}\frac{\Omega_{m}(a)}{a^{2}}\delta(a)=0,\label{delta22}
\eeq
with
\beq
\Omega_{m}(a)=\frac{\Omega_{m,0} a^{-3}}{E^{2}(a)}.
\eeq
Consequently, the growth of perturbations is governed by the background expansion through $H(a)$ and $\Omega_{m}(a)$, with the influence of different HDE models on structure formation arising solely through the expansion history. As a measure of structure formation, we introduce the growth factor $D(a)$
\beq
D(a)=\frac{\delta(a)}{\delta(a=1)},
\eeq
which is normalized to unity today, and the logarithmic growth rate
\beq
f(a)=\frac{d\ln D}{d\ln a},
\eeq
which, for smooth dark energy within general relativity, can be approximated as
\beq
f(a)\approx \Omega_{m}(a)^{\gamma},
\eeq
with $\gamma$ being the growth index. Then, the redshift--space distortion observable $f\sigma_{8}(z)$ is given by
\beq
f\sigma_{8}(z)=f(z)\sigma_{8}(z),
\eeq
where the root--mean--square amplitude of matter fluctuations at a scale of $8 h^{-1}$Mpc is taken as
\beq
\sigma_{8}(z)=\sigma_{8,0}\frac{D(z)}{D(0)},
\eeq
with $\sigma_{8,0}$ the value at the present epoch, calculated by CLASS using $H_{0}$ and $\Omega_{m,0}$ obtained from the MCMC samples, while the remaining cosmological parameters are fixed to their CLASS default values. This prescription amounts to predicting the growth amplitude for fixed early universe physics: the primordial scalar amplitude, the physical baryon and dark matter densities, the spectral index, and the radiation content are held at their default values, so that the model dependence of $\sigma_{8,0}$ enters only through $H_0$ and $\Omega_{m,0}$. Since the background datasets used here constrain mainly the late time expansion, we regard this as a consistent and minimal choice for a background conditioned comparison. It should be noted, however, that the absolute normalization of $\sigma_8$ and $S_8$ is not inferred jointly with the cosmological parameters, and the $S_8$ based comparison below should be interpreted in light of this external normalization choice.

\begin{figure*}[htp]
\begin{center}
\includegraphics[width=0.46\textwidth]{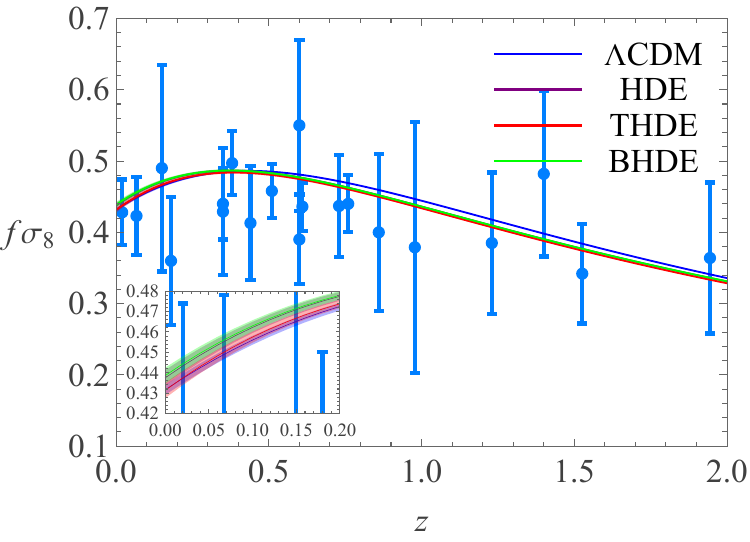}
\includegraphics[width=0.44\textwidth]{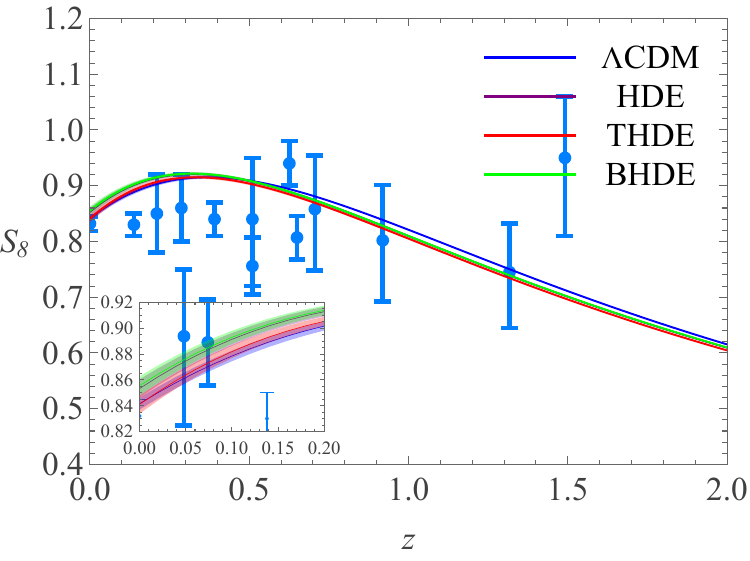}
\caption{\label{Fig5} Evolutionary curves of $f\sigma_{8}$ and $S_{8}$ for these HDE models with their mean values and 68\% CL bands.}
\end{center}
\end{figure*}

For these HDE models, the evolution of $H(a)$ and $\Omega_{m}$ can be obtained by solving the background dynamical Eqs.~(\ref{E22}) and~(\ref{Omde}). This expansion history then governs the growth of perturbations through Eq.~(\ref{delta22}). Using each set of parameters extracted from the MCMC samples, we solve the background and perturbation dynamical Eqs.~(\ref{E22}), ~(\ref{Omde}), and ~(\ref{delta22}), and display the resulting evolutionary curves of $f\sigma_{8}(z)$ with their mean values and 68\% CL bands in the left panel of Fig.~\ref{Fig5}. The independent $f\sigma_{8}(z)$ measurements used in this figure are taken from the compilation of~\cite{Kazantzidis2018} and further selected by~\cite{Das2026}, and are shown as blue error bars. These measurements directly probe the growth of cosmic structure via redshift--space distortions. As shown in this figure, the theoretical predictions for all models fall within the observational uncertainty bands across most of the redshift range, with only small deviations. Consequently, these models provide a fit to the data comparably well to the $\Lambda$CDM model. Specifically, the curves for the HDE and BHDE models overlap, the THDE model lies lowest, and at $z<0.5$, the THDE and $\Lambda$CDM models overlap.

Because the growth observables are computed from the background only posterior samples rather than sampled jointly with the growth data, the $\chi^2$ values below constitute a posterior predictive goodness of fit diagnostic, not a likelihood based model comparison. We therefore refrain from applying AIC or BIC to the growth sector, and we interpret the $\chi^2$ differences only as indicative trends. A consistent model selection statement combining background and growth data would require a joint likelihood, which is beyond the scope of the present work. To further quantify and investigate the discrepancies between theoretical predictions and observational data, we calculate the $\chi^{2}$ statistic
\beq
\chi^{2}=\sum\limits_{i}\frac{[f\sigma_{8}^{th}(z_i)-f\sigma_{8}^{obs}(z_i)]^{2}}{\sigma^{2}_{i}},
\eeq
where $f\sigma_{8}^{th}(z_i)$ represents the theoretical prediction, while $f\sigma_{8}^{obs}(z_i)$ denotes the corresponding observational measurement at $z_{i}$, and $\sigma_{i}$ is its uncertainty. The resulting $\chi^{2}$ values are summarized in Table~\ref{Tab3}. The THDE model yields the lowest $\chi^2$ value, with $\Delta\chi^2 = -1.439$ relative to $\Lambda$CDM; the HDE and BHDE models also yield slightly lower $\chi^2$ values than $\Lambda$CDM. These differences are not statistically significant, and we therefore do not interpret them as a preference. The appropriate conclusion is that current $f\sigma_8$ measurements are consistent with all four models within the adopted smooth dark energy approximation, and do not yet provide meaningful discrimination among them.

\begin{table}[h]
\centering
\caption{\label{Tab3} Comparison of the $\chi^2$ statistics for the $f\sigma_8$ and $S_8$ across these HDE models, evaluated as diagonal sums over the compiled measurements.}
\begin{tabular}{|c|c|c|c|c|}
\hline
\hline
\multirow{2}{*}{Model} & \multicolumn{2}{c|}{$f\sigma_{8}$} & \multicolumn{2}{c|}{$S_{8}$} \\
\cline{2-5}
 & $\chi^{2}$ & $\Delta\chi^{2}$ & $\chi^{2}$ & $\Delta\chi^{2}$ \\
\hline
$\Lambda$CDM & $10.785$ & $0$ & $54.576$ & $0$ \\
\hline
HDE & $10.164$ & $-0.621$ & $62.491$ & $7.915$ \\
\hline
THDE & $9.346$ & $-1.439$ & $54.775$ & $0.199$ \\
\hline
BHDE & $10.236$ & $-0.549$ & $63.785$ & $9.209$ \\
\hline
\hline
\end{tabular}
\end{table}

To further analyze the differences among these models, we calculate their $S_{8}(z)$ values using $S_{8}(z)=\sigma_{8}(z)\sqrt{\Omega_{m}(z)/0.3}$ and plot their evolutionary curves in the right panel of Fig.~\ref{Fig5}. The $S_{8}(z)$ measurements shown in this figure, adopted from~\cite{Atries2025, Qu2024, Adame2025} and further analyzed by~\cite{Akarsu2025}, are displayed as blue error bars, with the current value of $S_{8}$ taken from Planck 2018 results~\cite{Planck2020}. As shown in this figure, the theoretical predictions for all models fall within the observational uncertainty bands only at certain redshift points, while deviating from the data at other redshifts. These HDE models deviate slightly from the $\Lambda$CDM model, whereas HDE and BHDE models remain nearly indistinguishable from each other, and the THDE model lies lowest.

To quantify the discrepancies between theoretical predictions and observational data for $S_{8}$ in these models, we calculate the $\chi^{2}$ statistic
\beq\label{S8Chi2}
\chi^{2}=\sum\limits_{i}\frac{[S_{8}^{th}(z_i)-S_{8}^{obs}(z_i)]^{2}}{\sigma^{2}_{i}}.
\eeq
The corresponding $\chi^{2}$ values are presented in Table~\ref{Tab3}. The $S_8$ values show a noticeably larger spread among the models than the $f\sigma_8$ values: THDE lies close to $\Lambda$CDM ($\Delta\chi^2 = 0.199$), whereas HDE and BHDE yield $\Delta\chi^2 = 7.915$ and $9.209$, respectively. The compiled $S_8$ measurements~\cite{Akarsu2025} originate from different probes, including cluster abundances, CMB lensing, and spectroscopic clustering analyses~\cite{Atries2025, Qu2024, Adame2025}, with different fiducial cosmologies, redshift kernels, and nuisance treatments, and with no published covariance among them; the $\chi^{2}$ in Eq.~(\ref{S8Chi2}) is therefore a weighted least squares sum that treats these measurements as statistically independent. In addition, the absolute normalization of $\sigma_8$ follows an external prescription rather than being inferred jointly with the cosmological parameters, as discussed in Sec.~\ref{sec:4}. Consequently, the $\Delta\chi^2$ values for $S_8$ do not admit a formal model selection interpretation. They instead indicate a trend: within the present background conditioned and smooth dark energy framework, the predicted $S_8$ history of THDE remains close to that of $\Lambda$CDM, whereas those of HDE and BHDE lie above the bulk of the compiled measurements at $z\lesssim 1$. Whether this trend survives a joint background and growth inference with consistent growth measurements and full covariance remains an important question for future work.

\begin{figure*}[htp]
\begin{center}
\includegraphics[width=0.6\textwidth]{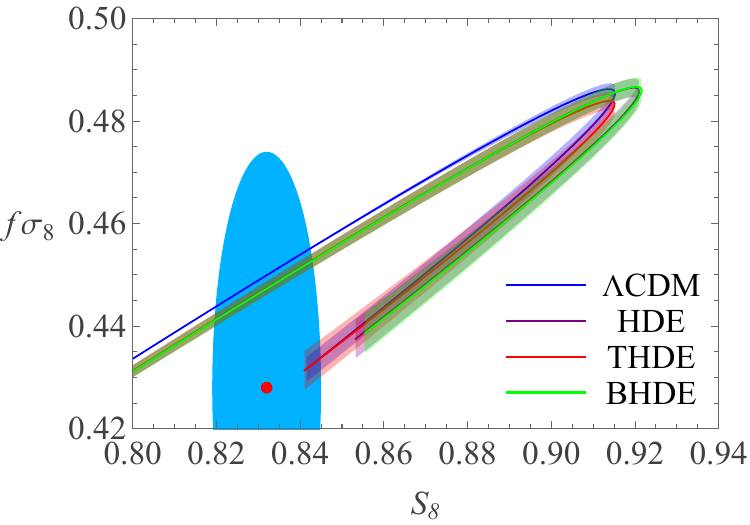}
\caption{\label{Fig6} Illustrative comparison of the predicted trajectories in the $(f\sigma_8, S_8)$ plane. The data region overlays an $f\sigma_8$ measurement at $z=0.020$~\cite{Kazantzidis2018} and the Planck 2018 $S_8$ value at $z=0$~\cite{Planck2020} without a known joint covariance; it is therefore intended as a qualitative overlay rather than a genuine two-dimensional likelihood contour.}
\end{center}
\end{figure*}

Among the three holographic models, THDE yields the lowest $\chi^2$ for both probes, although, as emphasized above, only the $f\sigma_8$ comparison admits a more straightforward statistical interpretation, and there the differences are not significant. However, despite the lower $\chi^{2}$ values of THDE, as shown in Fig.~\ref{Fig5}, the evolutionary curves of $f\sigma_{8}$ and $S_{8}$ for these HDE models exhibit similar behavior, with some curves even overlapping. To further investigate the differences in the evolutionary behavior of these HDE models, we construct the ($f\sigma_{8}$,$S_{8}$) plane, as shown in Fig.~\ref{Fig6}. In this figure, the red point represents the current value of ($f\sigma_{8}$,$S_{8}$), and the cyan region represents the $68\%$ CL. The current value of $S_{8}$ is taken from the Planck 2018 results, $S_{8}=0.832 \pm 0.013$~\cite{Planck2020}, while for $f\sigma_{8}$, which cannot be obtained from direct observations at $z=0$, we adopt the measurement at $z=0.020$ with $f\sigma_{8}=0.428 \pm 0.046$, as an approximation to the current value~\cite{Kazantzidis2018}. We stress that this plane is illustrative: the two coordinates originate from different measurements at slightly different redshifts ($z=0.020$ for $f\sigma_8$ and $z=0$ for $S_8$) without a known joint covariance, so the $68\%$ region does not represent a genuine joint likelihood contour. This figure shows that, in the ($f\sigma_{8}$,$S_{8}$) plane, the trajectories of the HDE and BHDE models overlap, and their current values do not fall into the $68\%$ CL region. Nevertheless, for the THDE and $\Lambda$CDM models, their trajectories overlap at late times and achieve the $68\%$ CL region of the current ($f\sigma_{8}$,$S_{8}$) data. Within this illustrative diagnostic, the THDE and $\Lambda$CDM trajectories reach the current data region, while those of HDE and BHDE do not; a quantitative use of this plane will require measurements of $f\sigma_8$ and $S_8$ at matched redshifts with known covariance.

In summary, the posterior predictive comparison shows that current $f\sigma_8$ measurements are consistent with all three holographic models and with $\Lambda$CDM, with no statistically significant discrimination. The $S_8$ comparison exhibits a larger spread: the predicted $S_8$ history of THDE remains close to $\Lambda$CDM, while those of HDE and BHDE lie above the bulk of the compiled measurements at $z\lesssim 1$, although the heterogeneity of the $S_8$ compilation and the external $\sigma_{8,0}$ normalization prevent a formal model selection interpretation. The illustrative $(f\sigma_8, S_8)$ plane shows the same trend. We therefore conclude that current structure formation data do not yet robustly distinguish among the HDE, THDE, and BHDE models, but that growth observables, $S_8$ type measurements in particular, offer the most promising avenue for future discrimination, to be realized with a joint background and growth likelihood and consistent growth measurements.

\section{Conclusions} \label{sec:5}

Based on the modified horizon entropy, different HDE models have been proposed. In this paper we have constrained the HDE, THDE, and BHDE models with the combined Pantheon+SH0ES SN Ia, OHD, DESI DR2, and Planck 2018 CMB distance prior data, and performed a posterior predictive comparison of the resulting linear growth histories with external $f\sigma_8$ and $S_8$ measurements.

For the background sector, with the Pantheon+SH0ES SN Ia, OHD, and DESI DR2 combination, the three holographic models yield marginally lower $\chi^2_{\min}$ values than $\Lambda$CDM but are increasingly penalized by AIC and BIC due to their extra parameters. Once the CMB distance priors are included, all three models are disfavored relative to $\Lambda$CDM according to $\Delta\chi^2_{\min}$, $\Delta$AIC, and $\Delta$BIC. The derived evolution of $\Omega_{\rm de}$ and $w_{\rm de}$ confirms that HDE and BHDE are nearly indistinguishable in these observables, while THDE separates from them only partially; background observables alone therefore cannot distinguish the three models.

For the growth sector, we solved the matter perturbation equation under the smooth dark energy approximation, in which the models differ only through their background expansion histories. The posterior predictive $\chi^2$ comparison shows that current $f\sigma_8$ measurements are consistent with all models, with no statistically significant discrimination ($\Delta\chi^2=-1.439$ for THDE relative to $\Lambda$CDM). The $S_8$ comparison shows a larger spread: the predicted $S_8$ history of THDE remains close to $\Lambda$CDM, whereas those of HDE and BHDE lie above the bulk of the compiled measurements $z\lesssim 1$. It should be emphasized, however, that the $S_8$ $\chi^2$ is a weighted least squares sum over heterogeneous measurements from different probes, with no common covariance published across them, and that the absolute normalization of $\sigma_8$ follows an external prescription rather than being jointly inferred; the $S_8$ differences should therefore be read as indicative trends rather than model selection criteria. The illustrative $(f\sigma_8, S_8)$ plane, constructed from measurements at slightly different redshifts, shows the same trend.

Our analysis indicates that current structure formation data do not yet robustly distinguish among the HDE, THDE, and BHDE models, but the $S_8$ sector exhibits the most promising discriminating trend. Realizing this potential requires a joint background and growth likelihood with a self-consistent treatment of the fluctuation amplitude, together with consistent growth measurements and published covariances; upcoming surveys such as DESI, Euclid, and LSST are expected to provide such data. Finally, our analysis neglects dark energy perturbations, so that the models differ in the growth sector only through their background histories; a complete treatment including dark energy perturbations and the nonlocal nature of the holographic IR cutoff may reveal genuinely perturbative differences among these models and is left for future work.

The results in ~\cite{Das2026} demonstrate that structure formation can be used to test the viability of the IR cutoff for HDE models, whereas our results point to a discriminating trend that may separate the THDE model from the HDE and BHDE models, suggesting that structure formation can serve as a useful probe for distinguishing HDE models. We note that only one observational data point is available in the $(f\sigma_{8},S_{8})$ plane of Fig.~\ref{Fig6}, whereas a comprehensive and quantitative diagnosis requires the values of $f\sigma_{8}$ and $S_{8}$ at the same redshift $z$ with higher precision. Future surveys such as DESI, Euclid, and LSST are expected to provide such high--precision measurements of $f\sigma_{8}$ and $S_{8}$ at matching redshifts, which will enable a more comprehensive diagnosis of these HDE models. It is also worth noting that the analysis in ~\cite{Das2026} and in this work is performed under the assumptions of negligible dark energy perturbations and the smooth dark energy approximation, in which the growth of perturbations is governed solely by the background expansion. When a non--minimal coupling is present in the dark sector, the perturbation of dark energy becomes non--negligible in overdense regions\cite{Saha2024, Saha2025}. A more complete treatment incorporating perturbations of dark energy and the nonlocal IR cutoff, or modifications to general relativity, could provide further discrimination among these HDE models.

\acknowledgments{This work was supported by the National Natural Science Foundation of China under Grant Nos. 12405081, 12305056, and 11865018. }

\end{document}